\documentclass[12pt,a4paper,psamsfonts]{article}
\usepackage{amsmath,amssymb,amsthm}
\usepackage{eucal}

\numberwithin{equation}{section}

\usepackage{geometry}
\def\a{\alpha}
\def\b{\beta}

\def\d{\delta}
\def\e{\epsilon}

\def\et{\eta}
\def\th{\theta}

\def\m{\mu}
\def\n{\nu}

\def\r{\rho}

\def\ph{\phi}

\def\ps{\psi}
\def\ome{\omega}

\def\G{\Gamma}
\def\D{\Delta}

\def\L{\Lambda}

\def\P{\Pi}

\def\P{\Phi}

\def\CC{\mathbb C}

\def\RR{\mathbb R}

\newcommand{\C}[1]{$(\ref{#1})$}

\def\o{\over}
\def\pa{\partial}

\makeatletter
\renewcommand\section{\@startsection {section}{1}{\z@}%
                                   {-3.5ex \@plus -1ex \@minus -.2ex}
                                   {2.3ex \@plus.2ex}%
                                   {\normalfont\large\bfseries}}
\renewcommand\subsection{\@startsection{subsection}{2}{\z@}%
                                     {-3.25ex\@plus -1ex \@minus -.2ex}%
                                     {1.5ex \@plus .2ex}%
                                     {\normalfont\bfseries}}
\makeatother

\begin{document}

\begin{center}
\addtolength{\baselineskip}{.5mm}
\thispagestyle{empty}
\begin{flushright}
{\sc MI-TH-1524}\\
\end{flushright}

\vspace{20mm}

{\Large  \bf M-theory and $G_2$ Manifolds}\footnote{Contribution to the ``{\it Focus Issue on Gravity, Supergravity and Fundamental Physics: the Richard Arnowitt Symposium}'', to be published in Physica Scripta. Based on a talk delivered by K. Becker at the workshop ``Superstring Perturbation Theory'' at the Perimeter Institute, April 22-24, 2015.
}
\\[15mm]
{Katrin Becker, Melanie Becker and Daniel Robbins}
\\[5mm]
{\it George P. and Cynthia W. Mitchell Institute for }\\
{\it Fundamental Physics and Astronomy, Texas A\& M University,}\\
{\it College Station, TX 77843-4242, USA}\\[5mm]

\vspace{20mm}

{\bf  Abstract}
\end{center}
In this talk we report on recent progress in describing compactifications of string theory and $M$-theory on $G_2$ and $Spin(7)$ manifolds. We include the infinite set of $\alpha'$-corrections and describe the entire tower of massless and massive Kaluza-Klein modes resulting from such compactifications.

\vfill
\newpage


\section{Introduction}

The equations of motion of eleven-dimensional supergravity are solved by a space-time of the form 
\begin{equation}
\RR^{1,3} \times M,
\end{equation}
{\it i.e.} a product of four dimensional Minkowski space and $M$, where $M$ is a compact seven-dimensional Ricci flat manifold. Moreover, the background four-form flux vanishes. Supersymmetry is unbroken if there is a parallel spinor along $M$.   In this case the metric on $M$ has $G_2$ holonomy. We are interested in full $G_2$ holonomy and not a proper subgroup.

Eleven-dimensional supergravity is the low-energy limit of $M$-theory. We expect an infinite series of corrections to the two-derivative action which are suppressed by the inverse radius of $M$. One can then ask: does the classical supergravity solution lift to a solution of $M$-theory?  This question can, of course, also be asked for supersymetric compactifications of string theory to a space-time of any dimension, {\it e.g.} $\RR^{1,2}\times M$. A related question is: can the classical metric of $G_2$ holonomy be modified to compensate for the $\a'$ corrections to the equations of motion and supersymmetry transformations? In string theory we have tools to explicitly compute $\a'$ corrections to {\it e.g.} Einstein's equations or supersymmetry transformations. 
However, computing all corrections is not feasible since there are infinitely many.
Luckily all the details of the $\a'$ corrected equations are not needed.
To explain this, we recall a classic theorem of mathematics. Let $M$ be a compact and oriented Riemannian manifold and let
\begin{equation}
H^p = \{ \ome \in E^p: \Delta \ome=0\},
\end{equation}
where $E^p$ is the space of smooth $p$-forms, be the harmonic forms. Here 
\begin{equation}
\D = d d^\dagger + d^\dagger d,
\end{equation}
is the Laplace-Beltrami operator. Then the following holds: 1) $H^p$ is finite dimensional and 2) there is the following orthogonal decomposition
\begin{equation}
E^p = \Delta(E^p)\oplus H^p.
\end{equation}
A consequence is that the equation $\D \ome=\a$ has a solution if and only if $\a$ is orthogonal to all harmonic forms. This is the Hodge decomposition theorem (see, {\it e.g.} 
ref. \cite{warner} for a proof and more details).
To solve the Laplace equation all we need to know is that the source is orthogonal to all harmonic forms but other details of the source are not needed.

This short note based on refs. 
\cite{Becker:2014rea}, \cite{Becker:2014uya}, \cite{Becker:2015wga} and  \cite{progress}
has two parts. In the first part we explain how the $\a'$ corrected supersymmetry constraints lead to equations of Laplace type. In the second part we explain how certain cohomology conditions needed for the solvability of the Laplace equation arise from space-time physics.

\section{Supersymmetry and the Laplace equation}

A basic tool to understand spaces on which parallel spinors exist are tensors constructed as spinor bilinears. In the present case such a bilinear takes the form
\begin{equation}
\phi_{abc} = \eta^T  \G_{abc} \eta,
\end{equation}
where $a,b,c$ are indices tangent to $M$ and $\G$ are the Dirac matrices. Moreover, the dual is
\begin{equation}
\psi= \star \phi.
\end{equation}
The three-form $\phi$ is in some sense more fundamental than the metric since given such a globally defined three-form a metric can always be constructed according to
\begin{equation}
g_{ab} = (\det s)^{-1/9} s_{ab},
\end{equation}
where
\begin{equation}
s_{ab} = - {1\o 144} \phi_{amn} \phi_{bpq} \phi_{rst} \e^{mnpqrst},
\end{equation}
and $\e=\pm 1$. If the spinor is covariantly constant

\begin{equation}
\nabla_a \eta=0,
\end{equation}
the classical solution emerges
\begin{equation}
d \phi = d \psi = 0 .
\end{equation}
Once $\a'$ corrections are included the gravitino supersymmetry transformation becomes
schematically
\begin{equation}\label{e27}
\d \psi_a = \nabla_a \et + A_a \eta + i B_a^b \G_b \eta,
\end{equation}
where the first term on the right hand side describes the classical contribution and the second and third terms encode the $\a'$ corrections. In seven dimensions irreducible spinors have eight real components and a basis of the space of spinors is given by $(\et, i \G_a \et)$ with $a = 1,\dots,7$. Equation \C{e27} describes an arbitrary spinor expanded into this basis. The coefficients of the expansion are tensors depending implicitly on the three-form
\begin{equation}
A_a = A_a[\phi], \qquad B_a^b = B_a^b [\phi].
\end{equation}
For example, to order $\a'^3$
\begin{equation}
A_a = 0 , \qquad B_a^b = \a'^3 \phi_{acd} \nabla^c Z^{db},
\end{equation}
where $Z$ is a tensor constructed from three Riemann tensors. But as previously mentioned, we do not need to know the concrete expressions for $A$ and $B$.

Our goal is to solve the spinor equations order by order in $\a'$. To order $\a'^3$ we find
\begin{equation}\label{ai}
\d \psi_a = \nabla_a' \et' + A_a \et + i B_a^b \G_b \et=0,
\end{equation}
where we used primes to denote the $\a'$ corrected quantities. Since $A$ and $B$ have an explicit overall $\a'^3$ they are constructed from the uncorrected three-form $\phi$.

Instead of solving eqn. \C{ai} directly it is more convenient (but equivalent) to solve the equations satisfied by the tensors
\begin{equation}
\phi_{abc}' = \et'^T \G'_{abc} \et', \qquad \ps' = \star' \ph'.
\end{equation}
Which equations determine $\phi'$ and $\psi'$? It turns out that we can get a complete set of equations by differentiating and anti-symmetrizing the previous expression. This leads to
\begin{equation}\label{aii}
\begin{split}
 d \ph' & = \a , \\
 d \ps' & = \b, \qquad \ps' = \star'  \phi'.
\end{split}
\end{equation}
The forms $\a$ and $\b$ can be written in terms of $A$ and $B$ using eqn. \C{ai}, see ref. \cite{Becker:2014rea} for details.

We can then ask: given $\a$ and $\b$,
does a solution of eqn. \C{aii} with a globally defined $\ph'$ and $\ps'$ exist? These equations are only solvable if $\a$ and $\b$ are exact. So we need
\begin{equation}
\begin{split}
& d \phi' = \a = d \chi, \\
& d \ps' = \b = d \xi.
\end{split}
\end{equation}
But why would $\a$ and $\b$ be exact? To order $\a'^3$ we have explicit expressions for $\a$ and $\b$ and we can verify that they are indeed exact. At order $\a'^n$, $n>3$ we shall use space-time physics to argue that $\a$ and $\b$ are exact, as described in detail in the next section. For now we
proceed assuming $\a$ and $\b$ to be exact.

The solution of the first equation in \C{aii} is then
\begin{equation}
\phi' =\phi + \chi + d b,
\end{equation}
where $b$ is a two-form. The equation determining $\ps'$ then becomes
\begin{equation}
d \psi' = d \star' \phi' = d \xi ,
\end{equation}
and by linearizing one obtains
\begin{equation}
\label{eq:MainG2Eq}
\star' \phi' = \star \phi + \star \left( {4\o 3} \pi_1 + \pi_7 - \pi_{27} \right)\d \phi.
\end{equation}
Here $\pi_{\bf{r}}$ is a projection of differential forms onto the subspace which transforms as a representation $\bf{r}$ of $G_2$.  For instance, for the three-forms we have a decomposition
\begin{equation}
\Lambda^3:=\Lambda^3T^*M=\Lambda^3_1\oplus\Lambda^3_7\oplus\Lambda^3_{27}.
\end{equation}
The condition (\ref{eq:MainG2Eq}) is a second order linear partial differential equation for $b$. Explicitly
\begin{equation}
d \star \left({4\o 3} \pi_1 + \pi_7 - \pi_{27}  \right)(\chi + d b) = d \xi.
\end{equation}
It turns out that this is a Laplace equation! To see this decompose
$b \in \L^2 = \L^2_7 \oplus \L^2_{14}$ and note that the $\L^2_7$ contribution corresponds to a diffeomorphism and can be discarded. We can then take\footnote{See ref. \cite{Becker:2014rea} for more details.}  $b\in \L_{14}^2$ and $d^\dagger b = 0 $ to obtain the partial differential equation
\begin{equation}
\D b = d^\dagger \rho,\qquad  b \in \L^2_{14}, \qquad d^\dagger b = 0 ,
\end{equation}
with
\begin{equation}
\rho = - \star \xi - (\pi_{27} - \pi_7 - {4\o 3} \pi_1 ) \chi.
\end{equation}
As discussed in ref. \cite{Becker:2014rea} the $d^\dagger\rho$ source satisfies $d^\dagger \rho \in \L_{14}^2$ and is orthogonal to the zero modes of the Laplacian (since it is co-exact).

This completes the existence proof and shows that to order $\a'^3$ there exists a background solving the $\a'$ corrected equations. What about the order $\a'^n$, $n>3$? As shown in ref. \cite{Becker:2014rea} a proof by induction works assuming $\a$ and $\b$ are exact order by order in $\a'$. So now the question remains: why are $\a$ and $\b$ exact to all orders in $\a'$? We address this question next.

\section{Space-time Physics}

In this section our goal is to show that the exactness of $\a$ and $\b$ is equivalent to the conditions for unbroken space-time supersymmetry. It is more convenient (but equivalent) to turn to $M$-theory at this point because the space-time physics arising from $M$-theory on $G_2$ manifolds is simpler than its string theory counterpart. Space-time fields arise from fluctuations around the background $\RR^{1,3} \times M$, where $M$ is a compact $G_2$ manifold.

There are infinitely many space-time fields with only a finite set corresponding to the massless fields and infinitely many massive modes. We use two guiding principles to construct the four-dimensional quantum field theory. First, supersymmetry in space-time. We assemble all fields into four-dimensional superfields, keeping supersymmetry in four dimensions manifest.
Second, we keep explicit locality along $\RR^{1,3} $ and $M$. Fields in four dimensions depend on the space-time coordinates. This is locality in four dimensions. Besides this we also keep locality along $M$ explicit. So fields are functions, forms or tensors on $M$, but we want to avoid doing any explicit spectral decomposition. We treat the coordinates on $M$ as continuous labels.

The eleven-dimensional fields (labeled by capital letters) decompose into several four-dimensional fields (labeled by greek letters)
\begin{equation}
\begin{split}
C_{MNP} & \to C_{abc}, C_{ab\m}, C_{a\m\n}, C_{\m\n\r}, \\
G_{MN} & \to g_{ab}, g_{a\m}, g_{\m\n}.
\end{split}
\end{equation}
Some fields are space-time scalars: $C_{abc}$ and $g_{ab}$, others are vectors, $C_{ab\m}$, $g_{a\m}$ and others anti-symmetric tensors of different types, $C_{a\m\n}$ and $C_{\m\n\r}$. There is also an infinite collection of spin 2 fields in space-time $g_{\m\n} = g_{\m\n}(x,y)$.

So far only pieces of the four-dimensional action with the above properties are known. For example, the action for all bosonic fields can be found in ref. \cite{Becker:2014uya} but the full action with manifest supersymmetry is work in progress \cite{progress}. However, what we know suggests that the cohomology conditions for $\a$ and $\b$ follow from this action as conditions for unbroken supersymmetry in four dimensions.

To explain the above statements we construct the most general action with the above properties including chiral and vector superfields. Moreover, we consider only global four dimensional supersymmetry.

We denote the coordinates on flat superspace in four dimensions by $x^\m$ and $\th$, the latter being a four component Majorana spinor. Superfields are then functions of these coordinates. So, for example, left-chiral or right-chiral superfields satisfy
\begin{equation}
{\cal D}_R \Phi(x,\th)=0, \qquad {\cal D}_L \P(x,\th)=0,
\end{equation}
respectively. Here ${\cal D}_{R,L}$ are the right-handed and left-handed components of the superderivative ${\cal D}$ respectively. Here and in the following we are using the conventions and notation of ref. \cite{Weinberg:2000cr}.
 There is an infinite family of left-chiral superfields
\begin{equation}
\P_{abc}(x,\th;y)= {\cal C}_{abc}(x;y)+\dots,
\end{equation}
which have discrete labels $a,b,c=1,\dots,7$ and a continuous label $y$, which is the coordinate on $M$. We denote the lowest component of the chiral superfield by ${\cal C}$ and the dots represent $\th$ dependent terms. ${\cal C}$ is a complex field and there is a natural choice
\begin{equation}
{\cal C}_{abc} = \hat \phi_{abc} +i C_{abc},
\end{equation}
where $C_{abc}$ is one of the components of the eleven-dimensional three-form and $\hat \phi_{abc}$ is related but not identical to the $G_2$-structure three form on $M$.

A supersymmetric action is constructed in the usual way
\begin{equation}
I = {1\o 2} \int d^4 x \left[ K(\P,\P^\dagger \right]_D + \int d^4 x \left[ f(\P) \right]_F + c.c.
\end{equation}
To illustrate the present framework we write the Lagrangian density for bosonic fields explicitly
\begin{equation}
\begin{split}
L = - \int d^7 y d^7 y' 
{\d^2 K \o \d {\cal C}(y)\d \bar {\cal C}(y')}
\left[ \pa_\m {\cal C}(y) \pa^\m \bar {\cal C}(y') - 
{\cal F}(y) {\cal F}^\star (y')\right]+
2 {\rm Re} \int d^7 y {\d f({\cal C}) \o \d {\cal C}(y)} {\cal F}(y),
\end{split}
\end{equation}
where we are suppressing the indices $a,b,c$. Here $\d$ is a functional derivative. The above expression will be local (it is equivalent to a single integral over $M$ of an expression built of local operators on $M$) if $K(\Phi,\Phi^\dagger)$ and $f(\Phi)$ are themselves local.  This follows since, for example, the second functional derivative of $K$ will produce a Dirac delta function $\d^7(y-y')$.

There is a natural choice for the superpotential that is local in this sense, namely
\begin{equation}
f (\P) \sim \int_M \P \wedge d \P,
\end{equation}
where $d$ is the exterior derivative along $M$. In a supersymmetric ground state
\begin{equation}
 {\d f \o \d \P} = 0,
\end{equation}
which implies $d \Phi=0$ or
\begin{equation}
\begin{split}
& d \hat \phi=0,\\
& G=0.
\end{split}
\end{equation}
This result is $\a'$ independent. To make contact to the previous discussion it is natural to identify
\begin{equation}
\hat \phi = \phi' - \chi.
\end{equation}
The exactness of $\alpha$ is then equivalent to the existence of the field redefinition between $\hat{\phi}$ and $\phi'$. To obtain more information we need to consider the K\"ahler potential described in terms of ${\cal C}$; these are coordinates of an infinite-dimensional complex manifold, which is the space of complex valued three-forms. $K$ defines a K\"ahler metric and the K\"ahler form on this space is
\begin{equation}
J = \int _{M \times M} d^7 y d^7 y ' {\d^2 K \o \d {\cal C}(y)\d\bar  {\cal C}(y')}
\d {\cal C}(y) \wedge \d \bar {\cal C} (y') = J_{A \bar B'} dy^A\wedge d \bar y ^{\bar B}.
\end{equation}
 We have introduced de-Witt notation in the second equality \cite{Becker:2014rea}. More explicitly, the index $A$ includes the index $a$ and the label $y$ and a sum over $A$ also includes an integral over $y$.
The K\"ahler metric has isometries. In particular, the shifts
\begin{equation}
{\cal C} \to {\cal C}' = {\cal C} + i d \L,
\end{equation}
leave the K\"ahler metric invariant. Since $J$ is a tensor
\begin{equation}
J \to J' = J + d (i_{d \L} J) + i_{d\L} (d J),
\end{equation}
where $i$ is the contraction with the vector field $d \L$. According to the Poincar\'e lemma there is a Killing potential $P_\L$ such that
\begin{equation}
i _{d \L} J = d P_{\L},
\end{equation}
where $P_{\L} $ is a map from the space of two-forms modulo closed two-forms into $\CC$, and $d$ on the right-hand-side is the exterior derivative on our infinite-dimensional K\"ahler manifold (as opposed to $d$ in the subscript on the left-hand-side, which is the exterior derivative on $M$).
We can then write
\begin{equation}
P_{ \L} = \langle \L, \m \rangle,
\end{equation}
where $\langle \cdot, \cdot \rangle$ is the inner product on forms and $\m$ is the moment map. This is a beautiful and extremely useful tool we borrow from symplectic geometry.  For us, the moment map is important since the space-time action includes a term of the form
\begin{equation}
I \sim \int D_{ab} \mu^{ab},
\end{equation}
where $D_{ab}$ is the auxiliary field of the vector multiplet containing $C_{ab\m}$. Integrating out $D_{ab}$ and requiring unbroken supersymmetry requires the moment map to vanish, which is equivalent to
\begin{equation}
\nabla_a \left( { \d K \o \d {\cal C}_{abc} } \right) =0.
\end{equation}
This is again valid to all orders in $\a'$ and it is equivalent to the existence of a co-exact two-form, or by Hodge duality, to an exact five-form. 
This is what we wanted to show. A similar analysis for compactifications on $Spin(7)$ manifolds can be found in ref. \cite{Becker:2014rea}. 

\section*{Acknowledgement}\addcontentsline{toc}{section}{Acknowledgement}

This work
was supported by the National Science Foundation grant PHY-1214344.


\providecommand{\href}[2]{#2}\begingroup\raggedright
\endgroup
\end{document}